\documentstyle[referee]{mn}
\font\bigmi = cmmi10 scaled \magstep2

\title[Limit-Cycle Behaviour of Thermally-Unstable Accretion Flows onto 
Black Holes]
{Limit-Cycle Behaviour of Thermally-Unstable Accretion Flows onto Black
Holes}
\author[Ewa Szuszkiewicz and John C. Miller]
{Ewa Szuszkiewicz$^{1,2}$ and John C. Miller$^{2,3}$\\
$^1$Astronomy Group, Department of Physics and Astronomy,
University of Leicester, University Road, 
Leicester $\,$LE1 7RH \\
$^2$International School for Advanced Studies, SISSA,
via Beirut 2-4, I-34013 Trieste, Italy \\
$^3$Nuclear and Astrophysics Laboratory, University of Oxford, 
Keble Road, Oxford $\,$OX1 3RH \\ }

\begin{document}

\maketitle 

\begin{abstract} 
\noindent 
Nonlinear time-dependent calculations are being carried out in order to
study the evolution of vertically-integrated models of
non-selfgravitating, transonic accretion discs around black holes. In this
paper we present results from a new calculation for a high-$\alpha$ model
similar to one studied previously by Honma, Matsumoto \& Kato who found
evidence for limit-cycle behaviour connected with thermal instability. Our
results are in substantial agreement with theirs but, in our calculation,
the disc material does not always remain completely optically thick and we
include a suitable treatment for this.  We followed the evolution for
several cycles and determined the period of the cycle as being about 780
seconds. Advective cooling is dominant in the region just behind the
outward-moving peak of surface density.  The behaviour of this model is
significantly different from what we saw earlier for low-$\alpha$ models
(which we discussed in a previous paper) and we contrast and compare the
two situations. 
\end{abstract} 

\section{Introduction}

Accretion discs can be subject to thermal instability (Pringle, Rees \&
Pacholczyk 1973) but occurrence of an instability does not necessarily
mean that after the characteristic growth-time, the disc will be
disrupted. The instability may also saturate in the non-linear regime
giving rise to continuing non-stationary behaviour of the system. This may
well be what is happening in the X-ray source GRS 1915+105 which has shown
a remarkable richness in its variability behaviour (Belloni et al. 1997).

In a previous paper (Szuszkiewicz \& Miller 1997 -- hereafter Paper I) we
reported on the first stages of a programme to investigate the evolution
of the thermal instability in vertically-integrated models of
non-selfgravitating, transonic accretion discs around black holes.  It was
found that for the original version of the slim-disc model (Abramowicz,
Czerny, Lasota \& Szuszkiewicz, 1988) with low viscosity ($\alpha$ =
0.001), the instability led to formation of a shock-like structure near to
the sonic point which, in the absence of special treatment, led to
termination of the calculation. This was very different from the behaviour
reported by Honma, Matsumoto \& Kato (1991, hereafter HMK) for a basically
similar model having the viscosity parameter $\alpha$ two orders of
magnitude larger. 

The aim of the present paper is to investigate a model with high $\alpha$,
similar to the one studied by HMK, following the same mathematical and
numerical approach as in Paper I, so as to clarify whether there is a
genuine difference between the low and high $\alpha$ regimes. The
parameter values used are the same as those chosen by HMK: 
$M=10M_{\odot}$, $\dot M = 0.06\,\dot M_{cr}$ and $\alpha = 0.1$ although
the resulting model is slightly different from theirs because of the
different mathematical formulation adopted. ($\dot M_{cr}$ is the critical
accretion rate corresponding to the Eddington luminosity.) We find
behaviour which is qualitatively similar to that presented by HMK and
decisively different from the low-$\alpha$ case. In our calculations we
obtained closed paths in the $\log T - \log \Sigma$ plane and followed the
evolution for three full cycles. In the short outburst phase the
assumption of the gas being optically thick is not fully satisfied and so
we have introduced an appropriate treatment for dealing with this and
recalculated the whole evolution. 
 
The plan of the paper is as follows. In Section 2 we summarize the basic
equations used in the calculations and introduce the treatment adopted for
dealing with the medium when it is not optically thick. In Section 3, we
present the global behaviour and phase space portraits as derived from our
calculations. The present results are compared with those of HMK in
Section 4 which also contains a discussion of the difference between the
low and high $\alpha$ cases. Section 5 is the conclusion. 

\section {Basic equations}

The equations used to describe the evolution of the thermal instability in
an axisymmetric, non-self-gravitating, optically-thick disc have been
presented in full detail in Paper I and here we just list them again for
completeness while describing in more detail the treatment which we have
used for the part of the flow which is not optically thick.

In cylindrical polar coordinates ($r$, $\varphi$, $z$) centered on the
black hole (having mass $M$), our basic hydrodynamical equations for
conservation of mass, energy and momentum are as follows: 
\begin{equation}
{D \Sigma \over D t} = - {\Sigma \over r} {\partial \over \partial r}
\left( r v_r \right)
\label{DSig}
\end{equation}

\begin{equation}
{Dv_r \over Dt} =
- {1 \over \rho  }{\partial p \over \partial r} +
 {{\left( l^2 - l_{_K}^2 \right) } \over r^3} 
\end{equation}

\begin{equation}
{Dl \over Dt}= -
{\alpha \over {r \Sigma}} {\partial \over \partial r} \left( r^2 pH 
\right)
\end{equation}

\begin{equation}
\Omega_{_K}^2 H^2 =6{p \over \rho} 
\end{equation}

\begin{equation}
\rho T {DS \over Dt} = Q_{vis} + Q_{rad} ,
\label{DS}
\end{equation}
Here, $D/Dt$ is the Lagrangian derivative given by
 
\begin{equation}
{D \over Dt} ={\partial \over \partial t}
+ v_r {\partial \over \partial r },
\end{equation}
$\Sigma=
\Sigma(r,t)$ is the surface density obtained by vertically integrating the
mass density $\rho$, $H$ is the half-thickness of the disc, 
$v_r = Dr/Dt $ (which is negative for an inflow),
$p$ is the pressure, $l=l(r,t)=rv_{\varphi}(r,t)$ is the specific
angular momentum, $l_{_K}$ is the value of $l$ for Keplerian motion
with $v_{\varphi}=
\left[ GMr/(r-r_{_G})^2 \right]^{1/2}$, $\Omega_{_K} = v_{\varphi}/r$,
$S$ is the entropy per unit mass, $T$ is the temperature, $Q_{vis}$
is the rate at which heat is generated by viscous friction
\begin{equation}
Q_{vis} = 
-\alpha pr\left({\partial \Omega_{_K} \over \partial r}\right) ,
\end{equation}
and $Q_{rad}$ is the
rate at which heat is lost or gained by means of radiative energy transfer
\begin{equation}
Q_{rad} = - {{F^-} \over {H}}
\end{equation}
with $F^-$ being the radiative flux away from the disc in the vertical
direction for which we use the expression
\begin{equation} 
F^- = {16 \sigma  T^4 \over  \kappa \rho H},
\label{Fmin}
\end{equation}  
where  $\sigma$ is the Stefan-Boltzmann constant and $\kappa$ is the
opacity. (Note that this expression is the same as that used for $F^-$ in
Paper I but written in terms of $\sigma$ rather than the radiation 
constant $a$.)

The thermodynamic quantities in the equatorial plane are taken to obey the
equation of state
 
\begin{equation}
p=k\rho T + {a \over 3}T^4
\end{equation}
and the opacity is approximated by the Kramers formula for chemical
abundances corresponding to those of Population I stars
   
\begin{equation}
\kappa = 0.34\, (1+ 6\times 10^{24}\rho T^{-3.5}) \ \ \
{\rm g}^{-1}\,{\rm cm}^2.
\label{Ross}
\end{equation}
If the medium is not optically thick, the diffusion approximation from
which the radiative flux equation (\ref{Fmin}) was derived is no longer
valid. Following Hur\'e et al (1994) (see also Lasota \& Pelat (1991);
Narayan \& Yi (1995) and Artemova et al. (1995)) we have then adopted a
general formula for $F^-$ based on the solution proposed by Hubeny (1990)
in the Eddington approximation: 

\begin{equation}
F^- = 6 { 4\sigma T^4 \over {{3\tau_{_R}  \over 2} + \sqrt{3}
+{1\over \tau_{_P} }}} 
\label{FHub}
\end{equation}
where $\tau_{_R}$ and $\tau_{_P}$ are the Rosseland and Planck mean
optical
depths (equal to $\kappa_{_R} \rho H$ and $\kappa_{_P} \rho H$).  The
expression
for the radiation pressure corresponding to this $F^-$ is 
\begin{equation} 
P_r = {F^- \over 12c} \left( \tau_{_R}  + {2\over \sqrt{3}}\right)
\label{PHub}
\end{equation}
 In the optically-thick limit $\tau_{_R}$ and $\tau_{_P}$ are much larger
than
unity and formulae (\ref{FHub}) and (\ref{PHub}) reduce to the standard
expressions for $F^-$ and $P_r$. The terms involving $\tau_{_R}$ are
dominant
here whereas in the opposite optically-thin limit, they are negligibly
small compared with other terms and $1/\tau_{_P}$ becomes the dominant
contribution to the denominator of (\ref{FHub}). Formulae (\ref{FHub}) and
(\ref{PHub}) provide a convenient interpolation between the two limits. 
The expressions used for $\tau_{_R}$ and $\tau_{_P}$ need to give
reasonable
results in the two limits. Here we adopt a very simple treatment. For the
Rosseland optical depth, we use the expression 
\begin{equation}
\tau_{_R} = 0.34 \, \Sigma \, (1+ 6\times 10^{24}\rho T^{-3.5}),
\end{equation}
corresponding to the $\kappa$ given by equation (\ref{Ross}), while for
the Planck optical depth we use
\begin{equation}
\tau_{_P} = {1\over 4\sigma T^4}(q^-_{br})
\end{equation} 
where $q^-_{br}$ is the bremsstrahlung cooling rate given by 
\begin{equation}
q^-_{br} =1.24\times 10^{21} H\rho^2T^{1/2} \ \ \
{\rm erg}\,{\rm cm}^{-2}\,{\rm s}^{-1}
\end{equation}
Of course, bremsstrahlung is only one of the important processes
contributing to the Planck opacity and it is our intention to include
others in future work. 

\section {\textfont1 = \bigmi
Results for the high--$\alpha$ model}

As mentioned in the Introduction, we are here focusing attention on a
particular model with black hole mass $M=10M_{\odot}$, initial accretion
rate $\dot M =0.06 \dot M_{cr}$ and viscosity parameter $\alpha = 0.1$. 
These parameters are the same as those used by HMK for their model having
the viscous stress component $\tau_{\varphi r}$ proportional to the total
pressure ($q = 0$ in their notation) but our model is not completely
identical to theirs because of the different formulations used. In this
section, we give a detailed presentation of the results which we have
obtained and delay until Section 4 comments on the comparison with those
of HMK.

Our computations were carried out using the Lagrangian hydrodynamics code
described in detail in Paper I. The grid organization and spacing were
similar to before but this time it was necessary to extend the grid out to
larger values of $r$ in order for conditions in the neighbourhood of the
outer boundary to remain essentially unchanging during the time of the
calculation, making it reasonable to impose constant state boundary
conditions there. (For safety, we put the outer boundary at $\sim 10^6
r_{_G}$, where $r_{_G}$ is the Schwarzschild radius $r_{_G}=2GM/c^2$.)

The initial model, constructed by solving the stationary version of
equations (\ref{DSig}) -- (\ref{DS}) for an optically thick medium, is
thermally unstable according to the local stability criterion with the
region of instability extending from $4.5\,r_{_G}$ to $17.5\,r_{_G}$.
(This is slightly different from the HMK model which was locally unstable
between $5.2\,r_{_G}$ and $13.1\,r_{_G}$.) Data from the stationary
calculation was then transferred onto the finite difference grid used for
the time-dependent calculation (as described in Paper I). Numerical noise
resulting from this and from the nature of the finite-difference treatment
was then sufficient to trigger growth of the instability. 

After less than one second of the evolution, two density waves are sent
out from around $6\,r_{_G}$ (within the locally-unstable region), one
moving inwards and the other outwards. This can be seen in Figure 1, where
the surface density $\Sigma$ is plotted against $r/r_{_G}$ at a succession
of times (marked in sequence from 1 to 11 with the dotted curve
corresponding to the beginning of the cycle).  The ingoing wave (see the
curve labelled 1) propagates quickly, passes through the sound horizon
(where the inward velocity is equal to the sound speed) and disappears
into the black hole, taking with it a significant amount of matter. This
is shown in more detail in Figure 2. The outgoing wave is launched on its
progress out through the disc only when the ingoing wave has fully passed
in through the sound horizon.  

At the onset of the instability, the temperature rises significantly in
the unstable region (see the upper panel of Figure 3), increasing the
contribution of the radiation pressure by nearly an order of magnitude. 
The first effect of this is to push the ingoing density wave into the
black hole leaving behind an underdensity while causing just a small
increase of density at the place from which the outgoing wave will be
launched. As the outgoing wave progresses outwards (with the Mach number
of the outflow $v_r/c_s$ reaching values as high as 0.2 -- see Figure 4)
the temperature peak is reduced but remains at a level still well above
that of the initial model. The outgoing wave heats the material through
which it passes, causing the part of the disc internal to it to swell up
as shown in Figure 5. Advective cooling is dominant in the region just
behind the peak of the wave. (Note that the sound speed $c_s$ used in
Figure 4 and elsewhere is the local {\it isothermal} sound speed
$(p/\rho)^{1/2}$ which is not exactly the quantity which is relevant for
considerations of causal connection. The effective ``sonic point'' for
these purposes comes {\it near} to the place where $v_r/c_s = - 1$ but
does not precisely coincide with it.)

In order for the outgoing wave to continue its propagation, it is
necessary that the material behind it should remain in a hot state. Once
the wavefront has moved beyond the linearly-unstable region, it becomes
progressively harder for the temperature to be maintained in the high
state and eventually the front starts to weaken and the temperature falls
dramatically down to a low state well below that of the initial model (see
curve 8). The front dies after about 20 seconds when it has reached about
$100\,r_{_G}$ and a slow filling up of the under-dense region then begins,
proceeding on the viscous timescale, with an associated progressive rise
in temperature.  Eventually, when the configuration has returned near to
its initial state, a second cycle begins which is very similar to the
first one. (There is a slight difference because the configuration does
not pass exactly through the stationary state at the end of the cycle.)

These results were obtained with formulae treating the medium as always
remaining optically thick. However, subsequent checking of the values of
the effective optical depth ($\tau_{eff} = (\tau_{_R}\tau_{_P})^{1/2}$)
revealed that this does not always remain much greater than unity when the
temperature is in the high state so that values for $F^-$ obtained using
expression (\ref{FHub}) deviate significantly from those obtained with
equation (\ref{Fmin}) which is appropriate for the optically-thick limit.
To check on the error introduced by this, we implemented the general
treatment described in Section 2 and repeated the calculation. The overall
behaviour was extremely similar and the only significant differences were
related to small changes in the temperature profiles in the inner parts of
the disc (see the lower panel of Figure 3). At the beginning of the second
and subsequent cycles, the ingoing wave developed a shock as it propagated
into the black hole (see Figure 6) whereas it remained regular when the
medium was treated as optically thick. Note that the shock is in the
supersonic part of the flow, inside the sound horizon. 

As in Paper I, it is valuable here also to consider a local view of the
results, taking cuts at particular values of $r$ and seeing how the values
of parameters there vary as the evolution proceeds. In particular, we
concentrate on evolution in the phase plane obtained by plotting $\log T$
against $\log \Sigma$. If one plots points corresponding to parameter
values at a specified value of $r$ for a sequence of stationary models
having the same $\alpha$ but progressively increasing values of the
accretion rate $\dot M$, then characteristic S-shaped curves are obtained.
In non-linear dynamics, existence of such S-shaped phase portraits of a
system is an indication of possible limit cycle behaviour. 

Figures 7, 8 and 9 show the phase portraits (upper panels) and temperature
time series at $5\,r_{_G}$, $10\,r_{_G}$ and $50\,r_{_G}$ respectively.
(There are only very small differences for these figures between the
optically thick calculation and the one with the general treatment -- the
results presented here are from the optically thick calculation.) The
location $5\,r_{_G}$ (Figure 7) lies just within the locally unstable
region for the initial model (in which parameter values lie on the middle
branch of the corresponding S-curves for stationary models) and the
initial state is represented by point 0. The S-curve for this value of $r$
is shown by the dotted line in Figure 7 (with the lower branch being
covered by computed points which we will now discuss). After the onset of
the instability, the subsequent behaviour follows a closed path in the
phase plane, passing successively through the points labelled 1, 2, 3,
etc., which correspond to the times for which the equivalently labelled
curves were drawn in the earlier figures. Each cycle lasts for
approximately 780 seconds and, as far as quantities at this location are
concerned, can be divided into two main sections. The first is the
outburst (for which a more detailed view of the temperature time series is
shown in Figure 10). This commences with the temperature rising rapidly
(points 1 and 2) and then falling again (point 3) while the surface
density decreases by more than an order of magnitude. This initial part
lasts for approximately one second. Following this, there is around 20
seconds of almost constant temperature during which $\Sigma$ first remains
essentially constant and then starts to gradually increase again with the
trajectory running parallel to the middle branch of the S-curve but not
quite lying on top of it. This gradual change is terminated by a dramatic
drop in temperature by a factor of three (point 8) down to the
low-temperature state. This marks the end of the first (outburst) stage
and the second (quiescent) stage then begins during which the evacuated
inner part of the disc is slowly refilled (on the viscous timescale). This
starts with a continuing slow decrease in temperature, lasting for around
50 seconds, with the trajectory following the lower branch of the S-curve.
$\Sigma$ is {\it decreasing} during this time but the decrease is then
reversed and the trajectory begins to move steadily up the lower branch
(with rising $T$ and $\Sigma$) for around 700 seconds until the turning
point is reached and another outburst begins. The calculation was
continued up to the beginning of the fourth cycle. The first cycle is
indicated with the solid line, the second with squares (marking the data
points plotted), the third with crosses and the fourth with triangles. The
later cycles are essentially identical to each other after a small change
from the first one.  Figures 8 and 9 show the equivalent phase portraits
and temperature time series for $10\,r_{_G}$ (within the locally-unstable
region at the initial time) and $50\,r_{_G}$ (which is outside it). 

\section {Discussion}

To make it easier to compare our results with those of HMK, the curves
shown in Figure 1 correspond to times chosen so as to have the front at
similar locations to those shown in Figure 5 of their paper. Our curves
labeled 1, 2, 3, 4, 5, 6 and 7 have the front at almost exactly the same
positions as for their curves with the same numbers but the front
propagates further out in our calculation and when it starts to break down
(curve 8), it has reached around $100\,r_{_G}$ rather than $70-80\,r_{_G}$
as in their calculations. Our curves 9 and 10 are at roughly the same
stages of the evolution as theirs. 

Our results are strikingly similar to those of HMK and, since the
numerical methods used in the two calculations were quite different, this
provides a convincing confirmation of the results obtained by HMK and of
the reality of the limit cycle behaviour. The main differences between our
results and theirs are in the distance to which the front propagates (as
mentioned above) and the amplitude of the density wave which is
considerably greater in our calculation. These differences are, at least
in part, due to the different mathematical formulation used (see Paper I
and our comment earlier in this paper about the size of the locally
unstable region) but they can also be related to differences in the
treatment of opacities and the diffusiveness of the numerical scheme (ours
has extremely low intrinsic numerical diffusion). Another difference is
that they find that the disc remains effectively optically thick
throughout whereas we do not, although the difference produced by this is
small. Also, we have explicitly shown that the behaviour repeats for a
number of cycles with those after the first one being essentially
identical. 

In Paper I, where we studied in detail a low-$\alpha$ model ($\alpha =
0.001$) having a locally unstable region of roughly the same extent as
that for the model treated in the present paper, we did not observe limit
cycle behaviour and the run ended with formation of a velocity spike
adjacent to the sonic point which appeared to be catastrophic. Now, using
similar numerical techniques and mathematical formulation, we have found
a completely different result for the high-$\alpha$ case and we need to
ask why this happens. A key difference is that for high $\alpha$, the
sonic point in the stationary models is outside the location of the
marginally stable orbit (at $3\,r_{_G}$) whereas for low $\alpha$, it is
inside it. Figure 11 shows the behaviour of $v_r/c_s$ plotted as a
function of $r/r_{_G}$ for the two cases at successive times during the
inward propagation of the disturbance produced when the thermal
instability is initiated. The top panel shows the high-$\alpha$ case and
the bottom panel is for low $\alpha$. Consider first the high-$\alpha$
case. The initial profile of $v_r/c_s$, shown by the upper dashed curve,
is distorted by the passage of the ingoing wave and progressively changes
into the qualitatively different profile shown by the lower dashed curve
which is then characteristic of the type of profile seen in the inner
parts of the disc during the outward propagation of the outgoing wave.
Note that the sonic point does not move inwards, remaining just external
to $3\,r_{_G}$, and that there is a striking pivot-like behaviour. (We
recall here our earlier comment that the effective sonic point for
considerations of causality does not precisely coincide with the point
where $v_r/c_s = - 1$ in these figures.) For low $\alpha$, the first of
the curves is the one shown with the dotted line and after a small initial
excursion in the outward direction, the sonic point is then pushed
substantially inwards with the profile steepening and then producing a
peak with positive (outward) velocities, leading to the formation of the
velocity spike shown in Figure 5 of Paper I which caused termination of
the run.  For high $\alpha$, the ingoing wave proceeds unimpeded into the
black hole and when this has happened, the outgoing wave is launched on
its progress out through the disc as described earlier. For low $\alpha$,
however, the ingoing wave was stopped in its inward progress by the
formation of the velocity spike and, while the beginnings of the outgoing
wave could be seen, the stage at which it could have been launched in its
outward progress through the disc was never reached. 

We have now made further refinements to the code which allow us to
continue the low-$\alpha$ calculation beyond the previous termination
point and it turns out that the growth of the velocity spike is not
catastrophic after all but rather that the amplitude subsides again after
reaching a maximum and the spike then keeps appearing and disappearing
during the subsequent evolution. Since the precise nature of this further
evolution seems to depend very much on how dissipative processes are
treated, we will delay detailed discussion of it until our next paper in
which we will study the effects of adopting a more physical viscosity
prescription than the simple $\alpha$ law used here. However, we should
report at this stage that by adding a large enough amount of artificial
diffusion of a particular type, we have found it possible to suppress the
effects of the velocity spike sufficiently so that an outward-going wave
{\it is} successfully launched out through the disc, giving a limit-cycle
behaviour very similar to that for the high-alpha case. We stress that
this is an artificially-produced result obtained by adding a dissipative
term in a way which we do not regard as being legitimate unless directly
motivated by a physical argument. We regard the account given in Paper I
as our definitive description of the low-$\alpha$ case within our present
physical assumptions and mathematical formulation and stress that the
appearance of the sonic-point velocity spike is part of a consistent
picture based on a fundamental underlying difference between the
low-$\alpha$ and high-$\alpha$ cases as discussed above. The calculation
leading to it has been extensively tested, as described in Paper I.  The
interest of our new experiment with the added artificial diffusion lies in
indicating that {\it if} real physical processes or improved mathematical
description were to produce a similar effect to that of the artificial
term which we introduced, then limit-cycle behaviour could still occur
also for the low-$\alpha$ case.  We will return to this discussion in the
Conclusion. 

Similar problems in understanding the physical nature of limit-cycle
behaviour have already been encountered in the context of Keplerian discs.
Taam \& Lin (1984) found a limit cycle connected with radiation pressure
driven thermal instability operating in Keplerian accretion discs around
black holes and neutron stars. Later, it was recognized by Lasota \& Pelat
(1991) that this was an artificially-produced result obtained because of a
particular treatment of the radiative cooling. They also pointed out that
the real nature of the limit cycle ought to be investigated outside the
thin disc approximation, as we are doing here, taking into account
departures from Keplerian motion and the transonic character of the flow.

\section {Conclusion }

In this paper, we have investigated the time evolution of a thermally
unstable transonic accretion disc with a high value for the viscosity
parameter $\alpha =0.1$.  We confirm the results obtained by Honma,
Matsumoto \& Kato (1991) who found evidence for limit-cycle behaviour for
a similar model.  We have continued our calculations for three complete
cycles and have found a period of about 780 seconds. 

The cycle can be divided into two distinct parts: an outburst phase and a
quiescent one. During the outburst phase, the flow ceases to be optically
thick in the inner part of the disc and we have implemented an appropriate
treatment for this regime. The results were compared with those from a run
in which the medium was treated as always being optically thick but only
small differences were found, probably because $\tau_{eff}$ never becomes
much smaller than unity. 

The non-stationary behaviour found for the high-$\alpha$ model is
significantly different from that for the low-$\alpha$ model presented in our
earlier paper (Szuszkiewicz \& Miller 1997), for which no limit-cycle
behaviour was seen. The key difference concerns what happens when the
ingoing wave reaches the sonic point.  For high $\alpha$, the location of
the sonic point remains essentially fixed and the ingoing wave passes
through it unimpeded and on into the black hole whereas for low $\alpha$
the sonic point was pushed inwards and a spike of positive velocity formed
just outside it, stopping the inflow. 

As outlined in Paper I, the strategy of the research programme, of which
the present paper forms a part, was to study first the simple case of the
original version of the slim-disc model, with all of its associated
assumptions and approximations, and then to consider the results obtained
for this as a standard reference against which to compare results from
subsequent more elaborate calculations in which additional effects would
be added one by one, thus giving a systematic way of understanding the
contribution from each of them. The present paper goes beyond Paper I in
considering a higher value of $\alpha$ but retains other simplifications
such as the $\alpha$ viscosity prescription, neglect of other
dissipative effects, assumption of hydrostatic equilibrium in the vertical
direction, use of the pseudo-Newtonian potential.  Our next step will
include introducing a more physical viscosity prescription with $\tau
_{\varphi r} \propto \partial \Omega / \partial r$ which makes the crucial
change of introducing a parabolic part into a system of equations which at
present is purely hyperbolic. The effect of such dissipative terms can
potentially be very important indeed. 

Our basic numerical scheme is extremely non-diffusive. While it normally
only requires the introduction of additional artificial viscosity in order
to deal with shocks, there can also be other circumstances in which it is
appropriate to introduce artificial diffusion, such as the one already
described in Paper I. However, it has to be said that this is a delicate
area forming part of the ``black art'' of numerical hydrodynamics which is
not often discussed in the literature. When is the adding of artificial
diffusion a benign thing, which allows one to smooth over shortcomings in
the numerical or mathematical representation of the problem, and when does
it become something drastic which changes the very nature of the problem?
When we first tried to run the high-$\alpha$ case presented in this paper,
we found that the outward-moving density wave was launched in exactly the
way described here (a quite different behaviour from the low-$\alpha$
case) but that numerical stability broke down as the wave died at the far
extent of its outward travel. We eventually cured this by introducing a
particular form of artificial diffusion based on a prescription kindly
supplied by F.~Honma (private communication).  Using an artificial term to
help an already-dying feature to go quietly seems quite legitimate in view
of the fact that it covers for the omission of real physical diffusive
effects which would be included in a full treatment.  On the other hand,
using heavy artificial diffusion to neutralise the effect of the
sonic-point velocity spike for low $\alpha$, does {\it not} seem
legitimate unless directly motivated by a physical argument. Maybe there
{\it is} such a motivation and maybe there {\it is} genuine limit-cycle
behaviour also for low $\alpha$ despite the fundamentally different
behaviour at the sonic point. However, this remains to be seen when
further relevant effects are included consistently in the calculations and
the next stage of our work will be pointed in that direction. 

\section* {Acknowledgments}
 
We gratefully acknowledge helpful discussions with Marek Abramowicz,
Bo\.{z}ena Czerny, Gabriele Ghisellini, Fumio Honma, Andrew King,
Jean-Pierre Lasota, Zbigniew Loska, Laura Maraschi, Derek Raine, Micha{\l}
R\'{o}\.{z}yczka, Marek Sikora, Roland Svensson and Andrzej Zdziarski. 
This work has been supported financially by the U.K. Particle Physics and
Astronomy Research Council and the Italian Ministero dell'Universit\`a e
della Ricerca Scientifica e Tecnologica.

\newpage
 
\section* {Figure Captions}

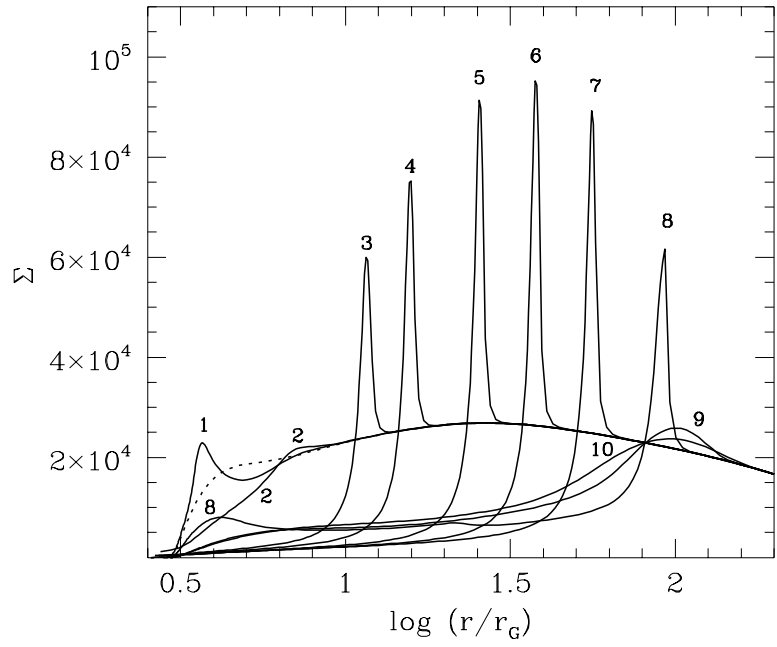
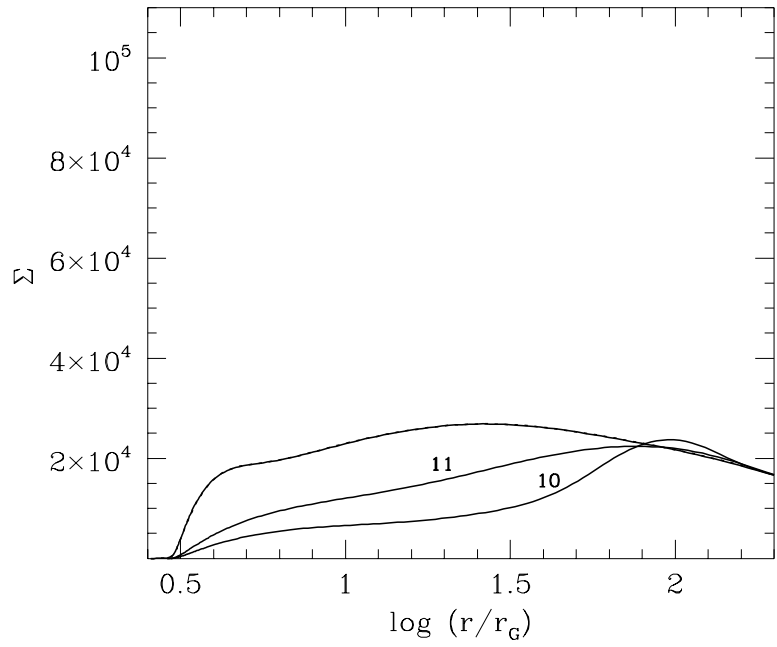
\noindent {\bf Figure 1:} Changes in the radial distribution of the
surface density $\Sigma$ (measured in units of g cm$^{-2}$) during the
second evolutionary cycle.  The curves marked 1 to 11 show the situation
after 876.41, 876.45, 876.79, 877.20, 878.36, 880.12, 883.42, 893.0,
925.0, 985.1 and 1209.0 seconds respectively. The dotted line corresponds
to the beginning of the cycle, at around 876 seconds, and the unmarked
solid line in the lower panel (which is essentially coincident with the
dotted one) is for the end of the cycle at 1657 seconds.

\noindent
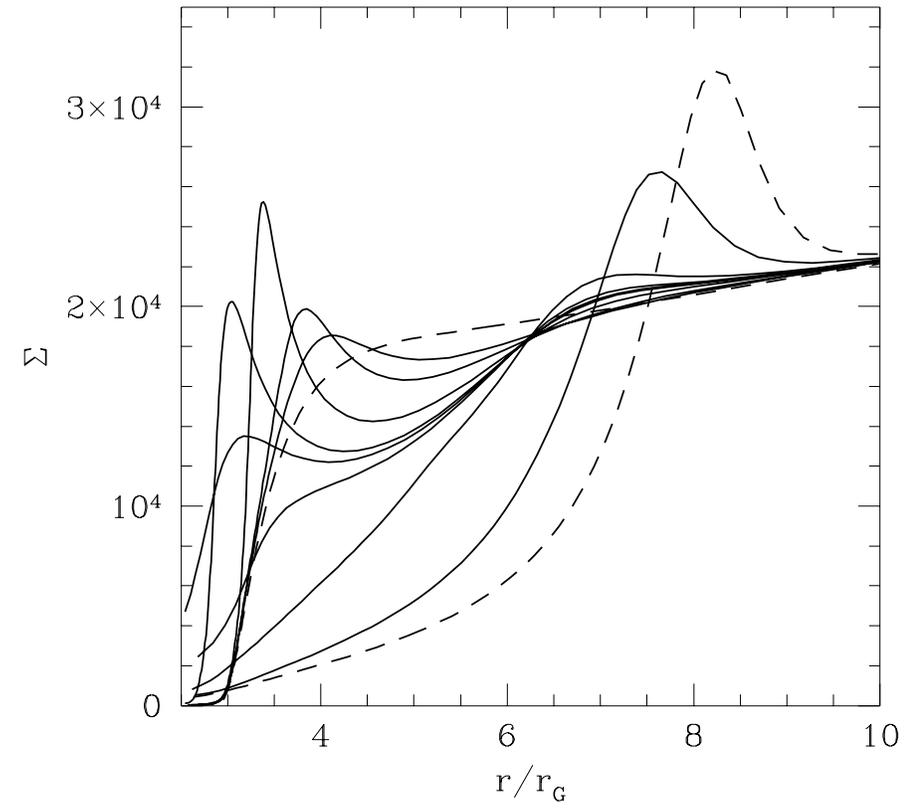
{\bf Figure 2:} 
A more detailed view of the onset of the instability, showing the
ingoing wave moving into the black hole and the subsequent launching
of the outgoing wave. The surface density is plotted at a succession of
times, with the dashed curves showing the profiles at the first and last
of these times (the first one being the upper curve at small $r$.) The
time elapsed between the first of the solid curves (when the perturbation
is first apparent) and the last of the solid curves (when the outgoing
wave is fully launched) is 0.14 seconds.

\noindent
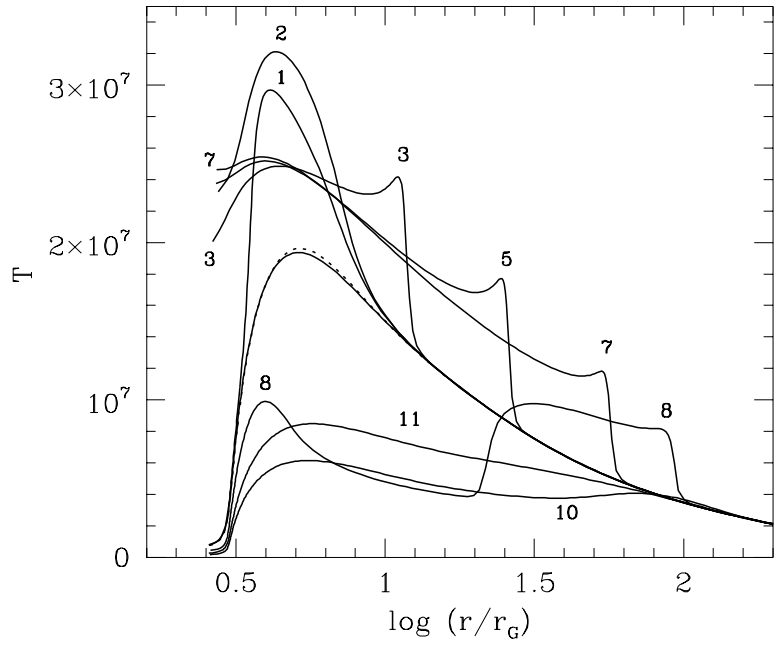
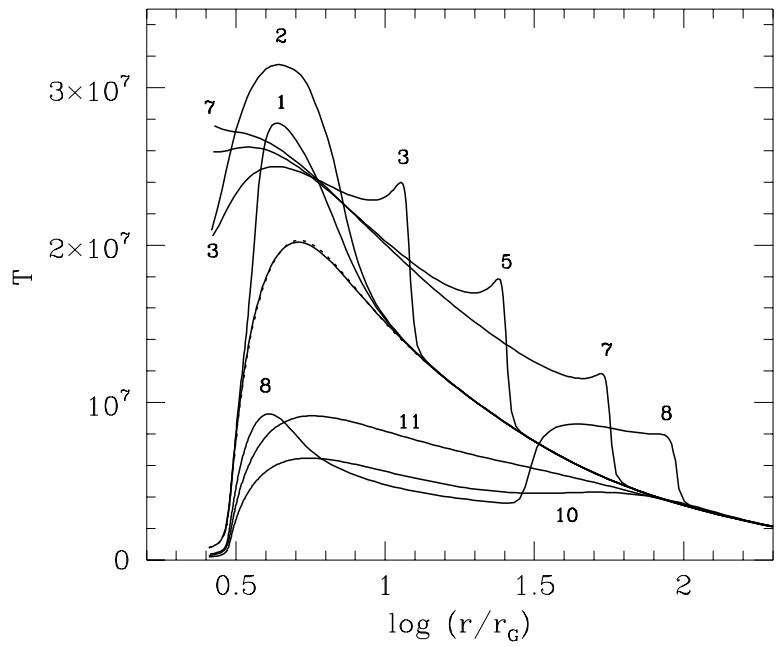
{\bf Figure 3:}
Evolution of the temperature profile in the calculation where the medium
is treated as always being optically thick (upper panel) and in the
calculation where regions which are not optically thick are treated in the
different way described in the text (lower panel). The temperature $T$ is
measured in degrees Kelvin and the labels correspond to the same times as
in Figure 1. 

\noindent
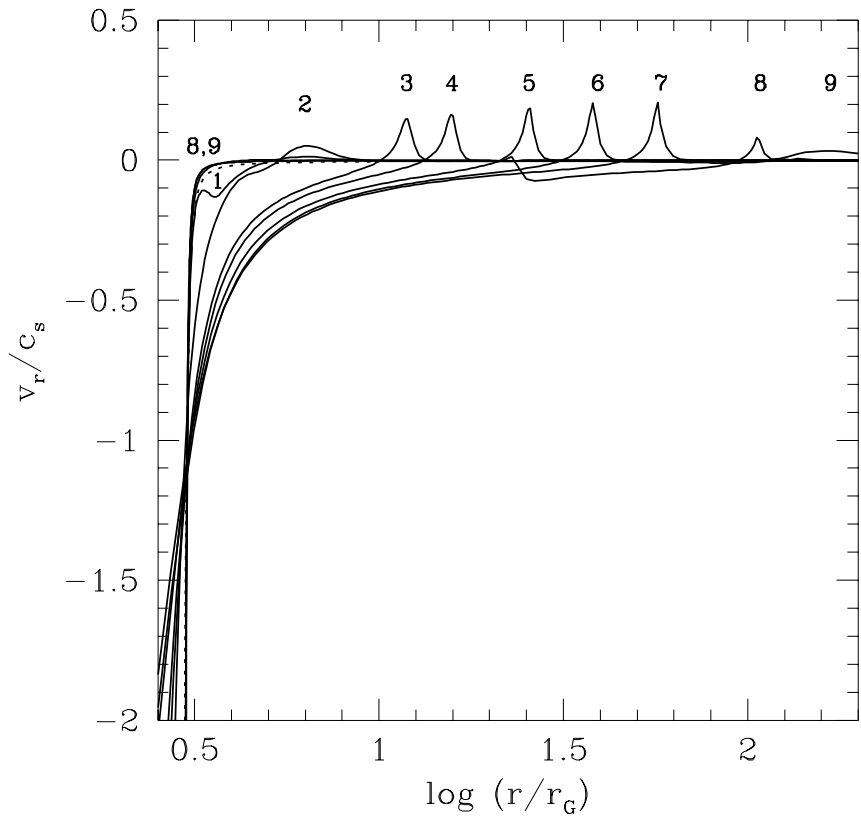
{\bf Figure 4:} 
The profile of the Mach number plotted for the same times as in Figure 1.
Note the marked change in the shape of these curves between the initial
time (marked with the dotted curve), for which there is a sharp ``elbow''
in the curve near to the sonic point, and the later times when there is a
gently changing slope behind the outgoing wave. 

\noindent
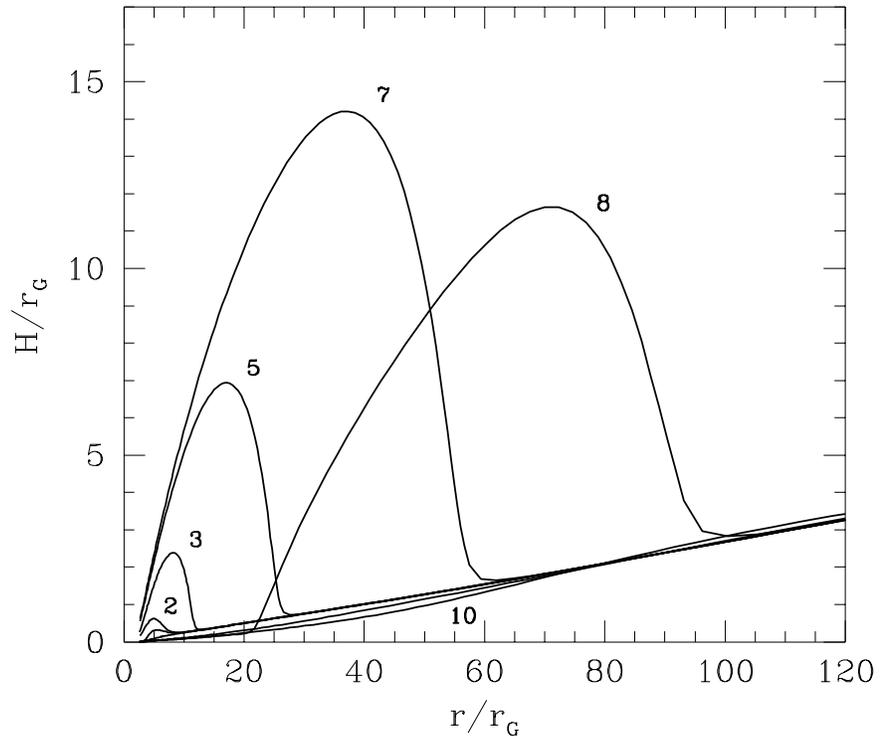
{\bf Figure 5:}
The half-thickness of the disc (in units of $r_{_G}$) plotted as a
function of $r/r_{_G}$ for some of the times shown in Figure 1.

\noindent
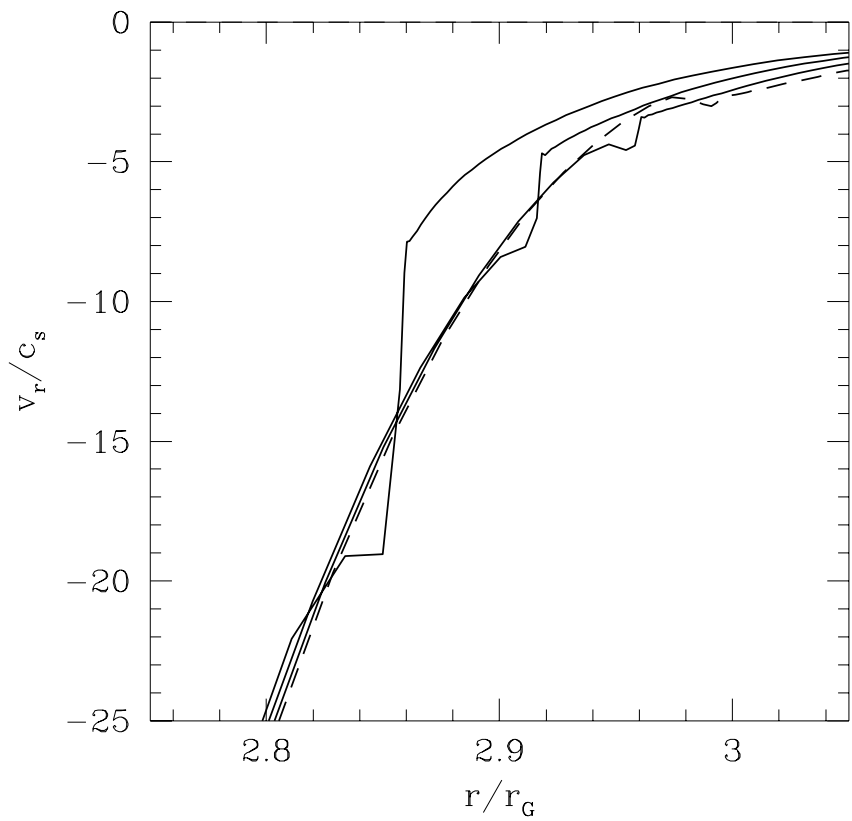
{\bf Figure 6:}
Formation of a shock in the ingoing wave at the beginning of the second
cycle in the calculation where the medium was not treated as if it were
always optically thick. The Mach number is plotted against $r/r_{_G}$
at a succession of times with the dashed curve being the first. The time
elapsed between the first and last curves is about 2 milliseconds. The
grid spacing seen in this very expanded picture may appear to be rather
coarse but we note that it was, in fact, quite adequate for following the
shock propagation shown which, in any case, is entirely inside the sound
horizon and hence does not affect the parts of the solution further out.

\noindent
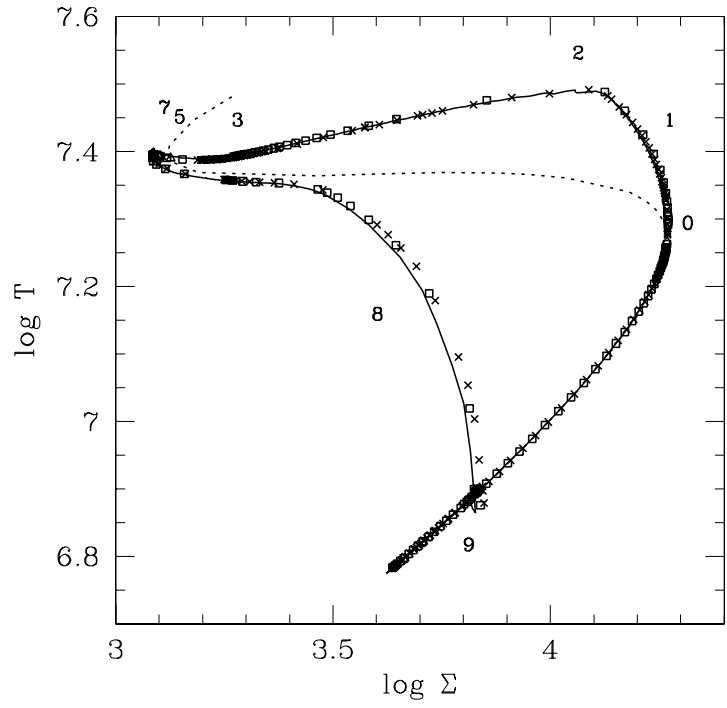
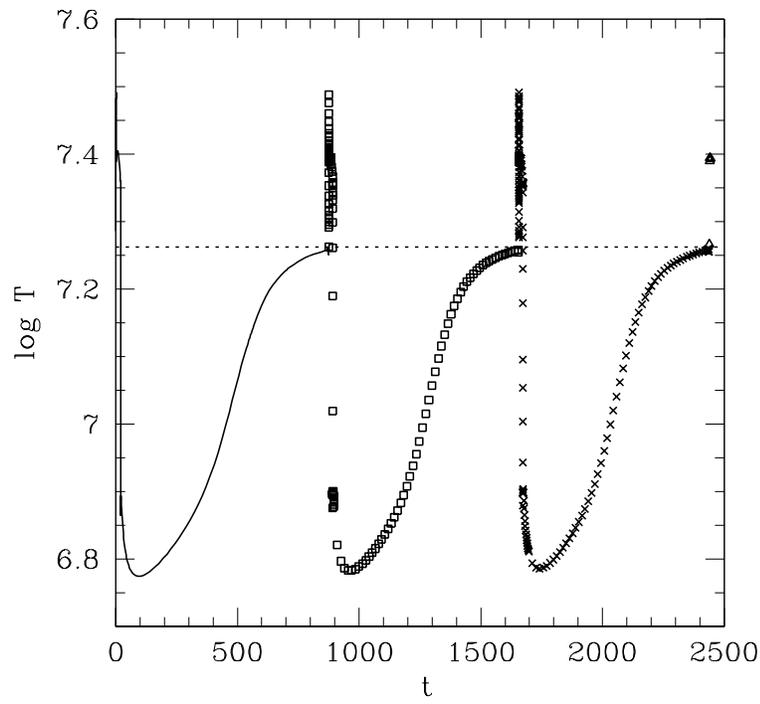
{\bf Figure 7:} 
The phase portrait in the $\log T - \log \Sigma$ plane (upper panel) and
the temperature time sequence (lower panel) for $5\,r_{_G}$. The time 
$t$ is measured in seconds.

\noindent
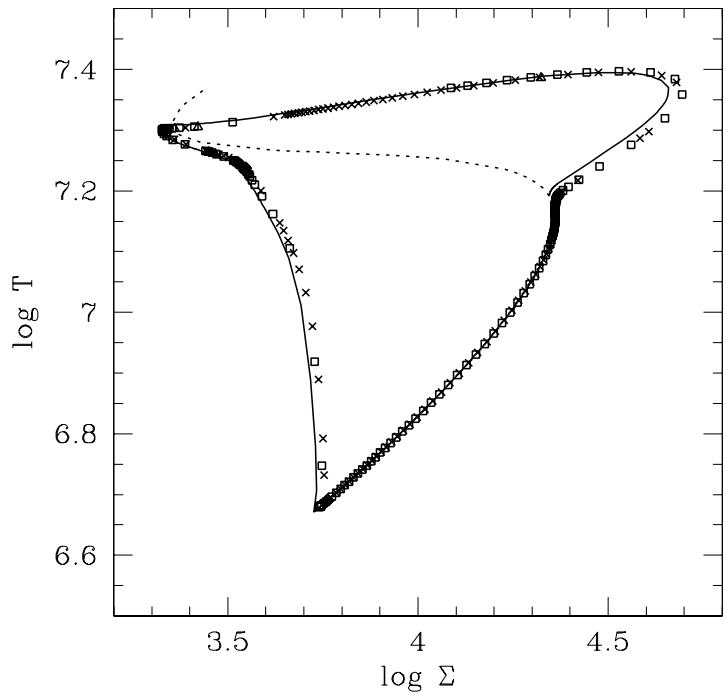
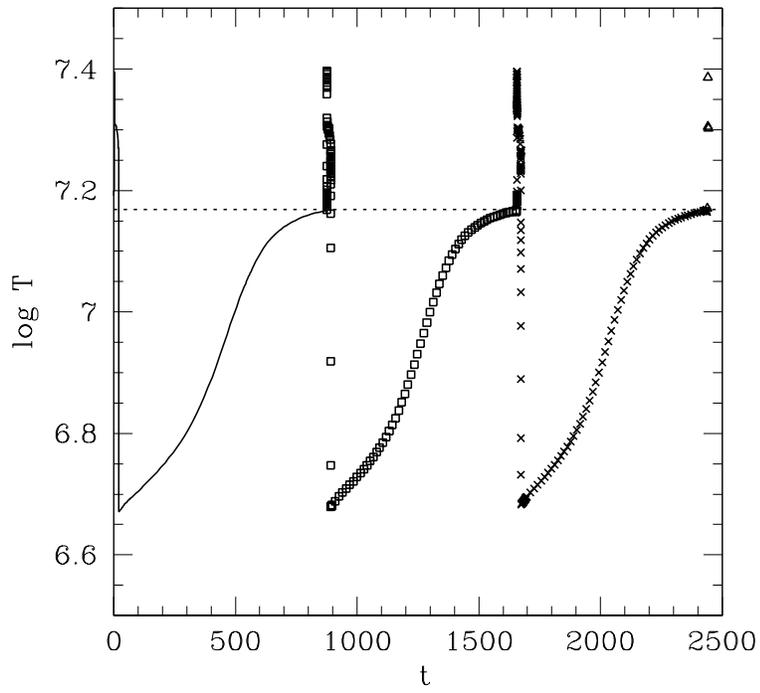
{\bf Figure 8:}
The phase portrait in the $\log T - \log \Sigma$ plane (upper panel) and
the temperature time sequence (lower panel) for $10\,r_{_G}$. 

\noindent
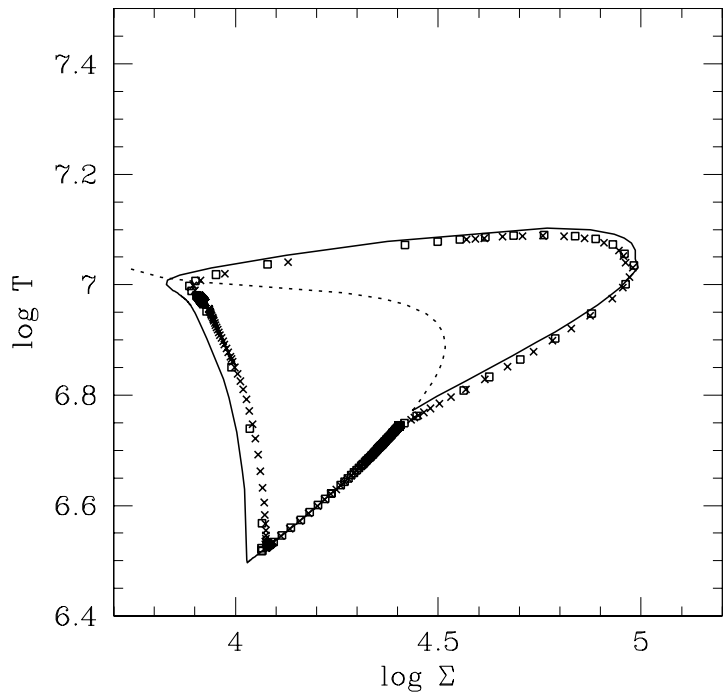
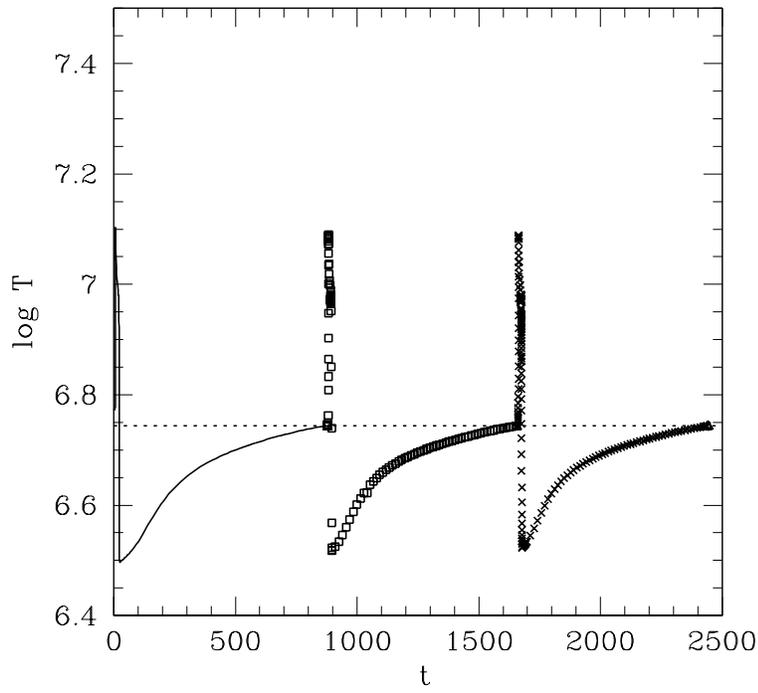
{\bf Figure 9:}
The phase portrait in the $\log T - \log \Sigma$ plane (upper panel) and
the temperature time sequence (lower panel) for $50\,r_{_G}$. 

\noindent
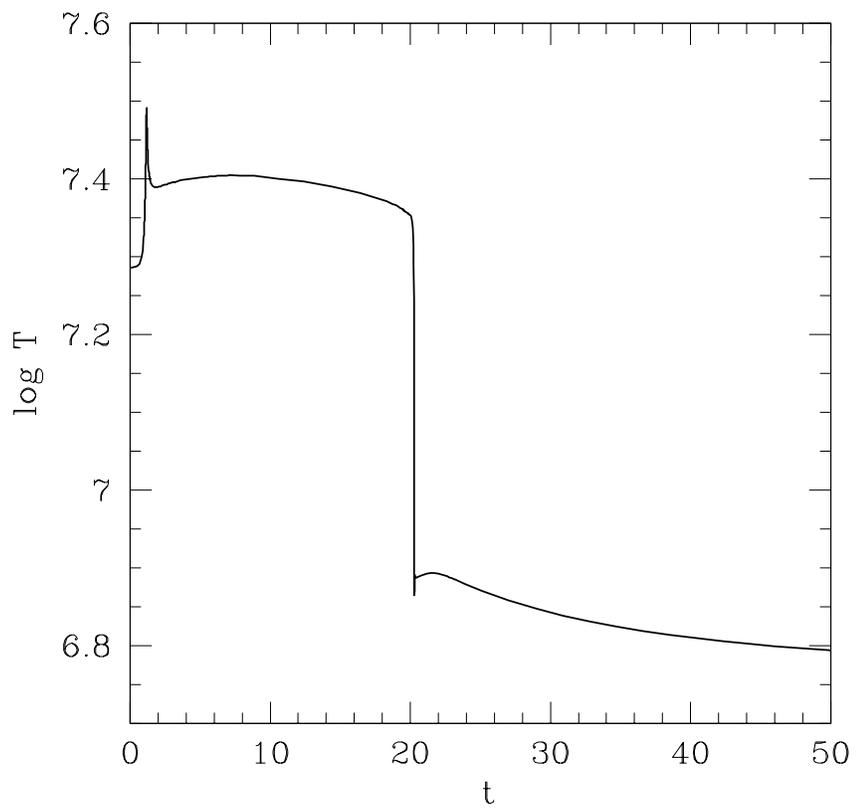
{\bf Figure 10:}
Expanded view of the temperature time sequence for $5\,r_{_G}$ during the 
initial outburst.

\noindent
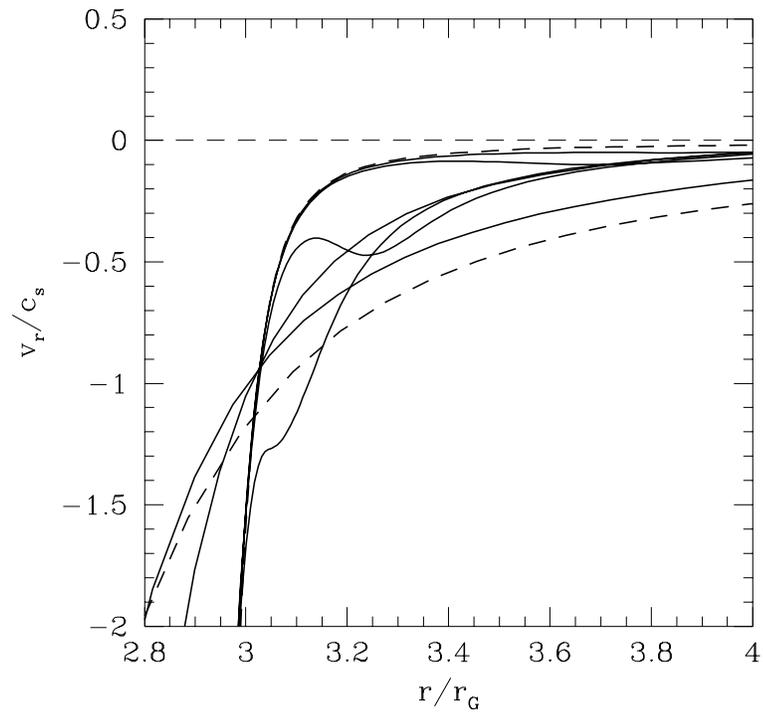
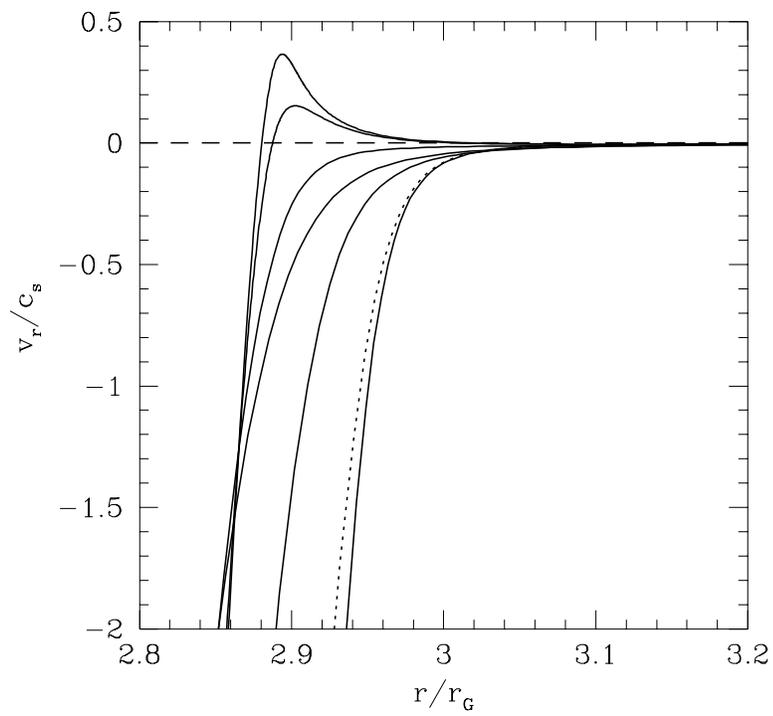
{\bf Figure 11:}
Comparison between the Mach number evolution for the high-$\alpha$ model
(upper panel) and the low-$\alpha$ model (lower panel). The elapsed time
is much greater for the low-$\alpha$ case than for the high-$\alpha$ one
($\sim 180$ seconds as compared with $\sim 1$ second). This is related to
the difference between the thermal timescales in the two cases.

\end{document}